\documentclass[reprint,
 amsmath,amssymb,
 aps,
]{revtex4-2}
\usepackage[colorlinks=true, linkcolor=blue, citecolor=blue, urlcolor=blue]{hyperref}
\usepackage{orcidlink}
\usepackage{graphicx}
\usepackage{dcolumn}
\usepackage{bm}
\usepackage{xcolor}
\begin{document}

\title{Kerr-Bertotti-Robinson Black Holes Surrounded by a Cloud of Strings}

\author{Faizuddin Ahmed\orcidlink{0000-0003-2196-9622}}
\email{faizuddinahmed15@gmail.com}
\affiliation{Department of Physics, The Assam Royal Global University, Guwahati 781035, Assam, India}

\author{\.{I}zzet Sakall{\i}\orcidlink{0000-0001-7827-9476}}
\email{izzet.sakalli@emu.edu.tr (Corresp. author)}
\affiliation{Department of Physics, Eastern Mediterranean University, Famagusta Northern Cyprus 99628, via Mersin 10, Turkiye}

\author{Ahmad Al-Badawi\orcidlink{0000-0002-3127-3453}}
\email{ahmadbadawi@ahu.edu.jo}
\affiliation{Department of Physics, Al-Hussein Bin Talal University, Ma'an 71111, Jordan}

\date{\today}

\begin{abstract}
In a recent study \cite{Podolsky2025}, authors introduced a new class of exact space-times in Einstein's gravity, which are Kerr black holes immersed in an external uniform magnetic field that is oriented along the rotational axis. Motivated by this work, we investigate a Kerr-like black hole solution with a cloud of strings surrounded by a uniform magnetic field. For the zero case, the space-time reduces to the Schwarzschild-Bertotti-Robinson black hole with a cloud of strings. Moreover, for zero magnetic field, the metrics simplify to a Kerr-like black hole surrounded by a cloud of strings, and its static counterpart reduces to the Schwarzschild black hole with a cloud of strings.
\end{abstract}

\maketitle

\small

\section{Introduction}\label{sec:1}

Black holes (BHs) constitute one of the most challenge predictions of Einstein's general relativity (GR) \cite{Misner1973,Carroll2004}, serving as natural laboratories for testing fundamental physics under extreme gravitational conditions. The Kerr solution \cite{Kerr1963}, describing a rotating BH characterized by mass $m$ and angular momentum parameter $a$, has become increasingly relevant following the historic detection of gravitational waves from binary BH mergers \cite{Abbott2016GW} and the first direct imaging of a BH shadow by the Event Horizon Telescope (EHT) collaboration \cite{Akiyama2019EHT,Akiyama2019}.

In string theory, the fundamental constituents of matter are extended one-dimensional objects rather than point-like particles \cite{Kiritsis2007}. Letelier \cite{Letelier1979} pioneered the investigation of gravitational fields sourced by a cloud of strings (CoS), deriving an exact solution for the Schwarzschild BH surrounded by strings within the framework of GR. This seminal work revealed that string matter distributions significantly modify spacetime geometry and BH properties. Rotating BHs are the most relevant subcases for astrophysics as it is believed that most astrophysical BHs are rotating. These solutions may also provide exterior metric for rotating stars. Subsequently, rotating BH solutions with CoS have been constructed and analyzed in various gravitational frameworks. Notably, Toledo et al. \cite{Toledo2020} applied the Newman-Janis algorithm \cite{Newman1965Kerr,Newman1965Rotating,AzregAinou2014} to obtain a Kerr-Newman-anti de Sitter black hole with quintessence and a spherically symmetric CoS. This algorithm provides a way to generate axisymmetric metrics from a spherically symmetric seed metric through a particular complexification of radial and (null) time coordinates, followed by a complex coordinate transformation. Often one performs eventually a change of coordinates to write the result in Boyer-Lindquist coordinates. Subsequently, following the similar approach, Li et al. \cite{Li2021} obtain a Kerr-like BH surrounded by CoS within Rastall gravity, investigating its weak gravitational lensing properties. More recently, Sun et al. \cite{Sun2024} explored the shadow of such BHs and constrained the string parameter using EHT observations of M87*. These comprehensive studies demonstrated that the string parameter substantially modifies the horizon structure, thermodynamic properties, photon sphere radius, innermost stable circular orbit (ISCO), and observational signatures compared to vacuum Kerr space-time.

Astrophysical BHs are typically embedded in external magnetic fields, which play crucial roles in accretion disk dynamics, jet formation mechanisms, and energy extraction processes \cite{Blandford1977,Meier2001,Janiuk2022,Ressler2025}. The presence of uniform magnetic fields aligned with the rotation axis introduces additional geometric structure to the spacetime. The Bertotti-Robinson (BR) spacetime, which describes the near-horizon geometry of extremally charged BHs and represents the direct product AdS$_2$ $\times$ S$^2$, provides a particularly elegant framework for studying electromagnetic field configurations \cite{Bertotti1959,Robinson1961,Cardoso:2004uz,Mazharimousavi:2009vh,Al-Badawi:2008ucc,badawi32}. In a groundbreaking recent study, Ref.~\cite{Podolsky2025} introduced a novel exact solution of Kerr BR (KBR) spacetime in the presence of a magnetic field. The line-element is described by

\begin{align}
ds^{2} &= \frac{1}{\Omega^{2}} \Big[ 
 -\frac{Q}{\rho^{2}} (dt - a \sin^{2}\theta d\varphi)^{2}
 + \frac{\rho^{2}}{Q} dr^{2} 
 + \frac{\rho^{2}}{P} d\theta^{2}\nonumber\\
 &+ \frac{P \sin^{2}\theta}{\rho^{2}} (a dt - (r^{2} + a^{2}) d\varphi)^{2}
\Big],\label{aa1}
\end{align}
where the metric functions are as follows:
\begin{align}
\rho^{2} &= r^{2} + a^{2} \cos^{2}\theta,\nonumber\\
P &= 1 + B^{2} \left( m^{2} \frac{I_{2}}{I_{1}^{2}} - a^{2} \right) \cos^{2}\theta, \nonumber\\
Q &= (1 + B^{2} r^{2}) \Delta,\nonumber\\
\Omega^{2} &= (1 + B^{2} r^{2}) - B^{2} \Delta \cos^{2}\theta,\nonumber\\
\Delta &= \left( 1 - B^{2} m^{2} \frac{I_{2}}{I^2_{1}} \right) r^{2} - 2 m \frac{I_{2}}{I_{1}} r + a^{2},\nonumber\\
I_{1} &= 1 - \tfrac{1}{2} B^{2} a^{2}, \quad
I_{2} = 1 - B^{2} a^{2}. \label{aa2}
\end{align}

This solution generalizes the classical Kerr geometry by incorporating a uniform magnetic field strength $B$ oriented along the rotational axis. The BR electromagnetic field configuration \cite{Bertotti1959,Robinson1961,Asher1960} represents a special case where the electromagnetic stress-energy tensor mimics a cosmological constant, yielding remarkable exact solutions. The metric exhibits intriguing features including modified horizon structure, ergosphere, and photon regions due to magnetic field effects \cite{Zeng2025}. Importantly, this solution admits well-defined limiting cases:

\begin{itemize}
    \item For $B=0$: one can recover Kerr-like solution as,
    \begin{align}
    ds^{2} &= 
 -\frac{\Delta}{\rho^{2}} (dt - a \sin^{2}\theta d\varphi)^{2}
 + \frac{\rho^{2}}{\Delta} dr^{2} 
 + \rho^{2} d\theta^{2}\nonumber\\
 &+ \frac{\sin^{2}\theta}{\rho^{2}} [a dt - (r^{2} + a^{2}) d\varphi]^{2},\label{aa3}
\end{align}
where
\begin{equation}
    \rho^{2}= r^{2} + a^{2} \cos^{2}\theta,\quad \Delta=r^{2} - 2 m r + a^{2}.\label{aa4}
\end{equation}

\item For $a=0$: one recovers a Schwarzschild-BR-like metric as follows:
\begin{align}
ds^{2}= \frac{1}{\Omega^{2}} \Big[
 -\mathcal{Q}\,dt^{2}
 + \frac{dr^2}{\mathcal{Q}}
 + \frac{r^{2}}{P}\,d\theta^{2}+ P\,r^{2}\sin^{2}\theta\,d\phi^{2}
\Big],\label{aa5}
\end{align}
where the metric functions are given by
\begin{align}
P &= 1 + B^{2}m^{2}\cos^{2}\theta,\nonumber\\
\mathcal{Q} &= (1 + B^{2}r^{2})\left[1- B^{2}m^{2}- \frac{2m}{r}\right], \nonumber\\
\Omega^{2} &=1 + B^{2}\left[r^{2} \sin^2 \theta+ ( B^{2}m^{2} r^{2} + 2mr)\cos^{2}\theta\right]. \label{aa6}
\end{align}
\end{itemize}

Inspired by the  string matter distributions and external magnetic fields, a fundamental question naturally emerges: What are the combined effects of a CoS and a uniform BR magnetic field on rotating BH spacetimes? This question is particularly compelling for understanding BH physics in string-theoretic scenarios where both string matter and magnetic fields may coexist. The interplay between the string parameter and the magnetic field strength can potentially lead to novel signatures in BH thermodynamics, horizon structure, geodesic motion, and observational properties.

In this work, we present an exact solution describing a KBR BH surrounded by a CoS and immersed in a uniform BR magnetic field. This solution extends the KBR spacetime \cite{Podolsky2025} by including the string cloud as an additional matter source. In particular, it reduces to the Letelier-Kerr BH in the absence of a BR magnetic field, to the Letelier-BR BH for a non-rotating BH, and to the original Letelier solution \cite{Letelier1979} for a non-rotating BH without a BR magnetic field. We investigate how the string cloud influences the horizon structure, thermodynamic quantities (entropy, temperature, surface gravity), photon sphere, ISCO in the magnetized environment. Our analysis shows several novel features and highlights discrepancies with previous studies \cite{Podolsky2025,Zeng2025}, particularly concerning horizon locations and surface gravity.

\section{KBR BH surrounded by C\lowercase{o}S}\label{sec:2}

Inspired by the metric (\ref{aa1}), we consider the following rotating BH solution surrounded by a CoS immersed in a uniform magnetic field described by  
\begin{align}
ds^{2} &= \frac{1}{\Omega^{2}} \Big[ 
 -\frac{Q}{\rho^{2}} (dt - a \sin^{2}\theta d\varphi)^{2}
 + \frac{\rho^{2}}{Q} dr^{2} 
 + \frac{\rho^{2}}{P} d\theta^{2}\nonumber\\
 &+ \frac{P \sin^{2}\theta}{\rho^{2}} (a dt - (r^{2} + a^{2}) d\varphi)^{2}
\Big],\label{bb1}
\end{align}
where the metric functions are as follows:
\begin{align}
\rho^{2} &= r^{2} + a^{2} \cos^{2}\theta,\nonumber\\
P &= 1 + B^{2} \left( m^{2} \frac{I_{2}}{I_{1}^{2}} - a^{2} \right) \cos^{2}\theta, \nonumber\\
Q &= (1 + B^{2} r^{2}) \Delta,\nonumber\\
\Omega^{2} &= (1 + B^{2} r^{2}) - B^{2} \Delta \cos^{2}\theta,\nonumber\\
\Delta &= \left( 1-\alpha - B^{2} m^{2} \frac{I_{2}}{I^2_{1}} \right) r^{2} - 2 m \frac{I_{2}}{I_{1}} r + a^{2},\nonumber\\
I_{1} &= 1 - \tfrac{1}{2} B^{2} a^{2}, \quad
I_{2} = 1 - B^{2} a^{2}. \label{bb2}
\end{align}
Its static counter part (Schwarzschild-BR-like surrounded by a cloud of strings ) is given by 
\begin{align}
ds^{2}= \frac{1}{\Omega^{2}} \Big[
 -\mathcal{F}\,dt^{2}
 + \frac{dr^2}{\mathcal{F}}
 + \frac{r^{2}}{P}\,d\theta^{2}+ P\,r^{2}\sin^{2}\theta\,d\phi^{2}
\Big],\label{bb3}
\end{align}
where the metric functions are given by
\begin{align}
P &= 1 + B^{2}m^{2}\cos^{2}\theta,\nonumber\\
\mathcal{F} &= (1 + B^{2}r^{2})\left[1-\alpha- B^{2}m^{2}- \frac{2m}{r}\right], \nonumber\\
\Omega^{2} &=1 + B^{2}\left[r^{2} \sin^2 \theta+ ( B^{2}m^{2} r^{2} + 2mr)\cos^{2}\theta\right]. \label{bb4}
\end{align}

For zero magnetic field, $B=0$, the metric (\ref{bb1}) simplifies to the Kerr-like BH solution with a CoS \cite{Li2021} in the framework of Einstein gravity given by
\begin{align}
    ds^{2} &= 
 -\frac{\Sigma}{\rho^{2}} (dt - a \sin^{2}\theta d\varphi)^{2}
 + \frac{\rho^{2}}{\Sigma} dr^{2} 
 + \rho^{2} d\theta^{2}\nonumber\\
 &+ \frac{\sin^{2}\theta}{\rho^{2}} [a dt - (r^{2} + a^{2}) d\varphi]^{2},\nonumber\\
 \rho^{2}&= r^{2} + a^{2} \cos^{2}\theta,\quad \Sigma=(1-\alpha) r^{2} - 2 m r + a^{2},\label{bb5}
\end{align}
whereas the corresponding static metric (\ref{bb3}) reduces to the Letelier BH solution \cite{Letelier1979}. 

\begin{figure}[ht!]
    \centering
    \includegraphics[width=0.8\linewidth]{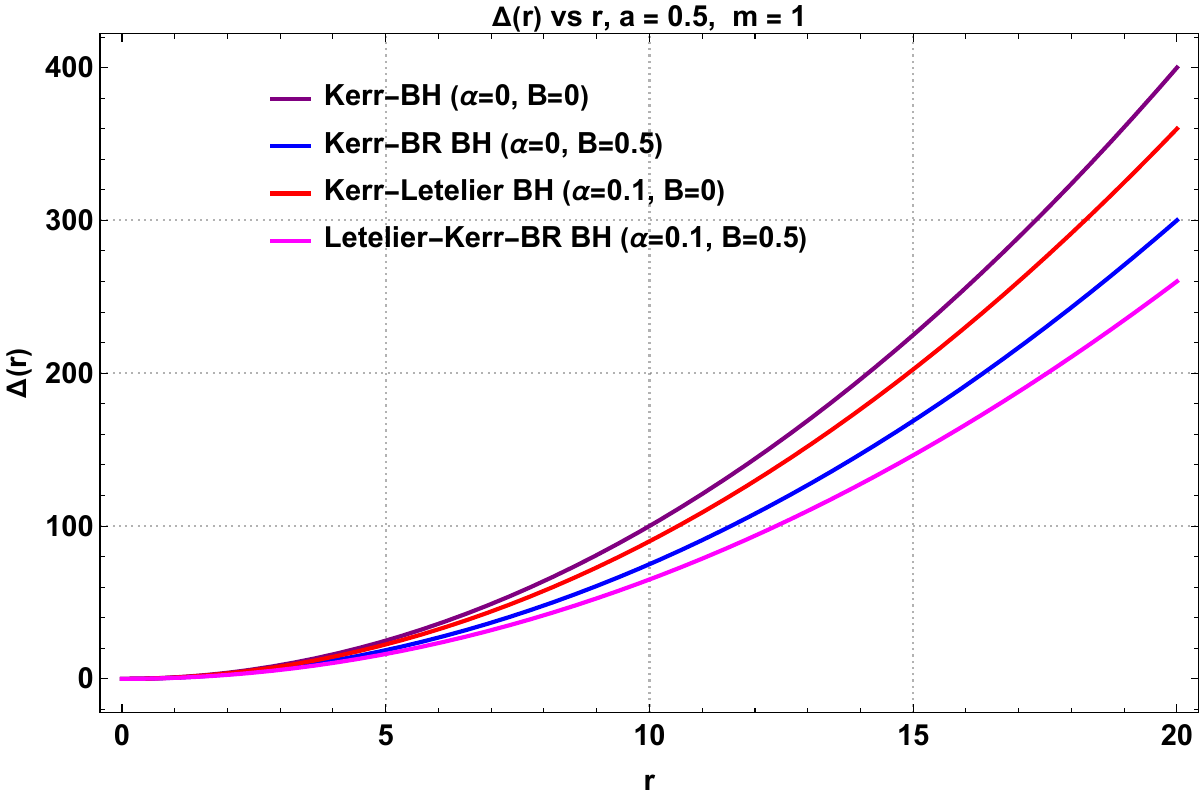}\\
    \includegraphics[width=0.8\linewidth]{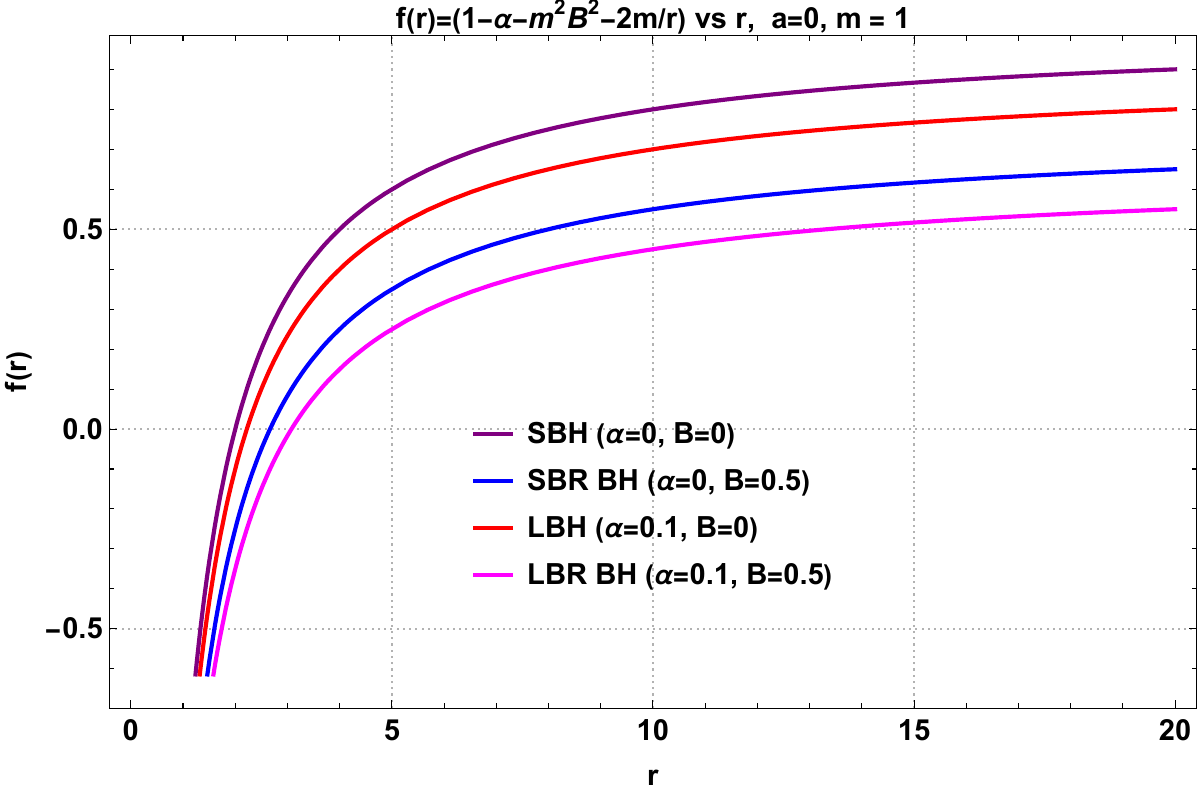}
    \caption{\footnotesize Variation of $\Delta$ (rotating BH) and $f(r)=\left(1-\alpha- B^{2}m^{2}- \frac{2m}{r}\right)$ (non-rotating BH) as a function of the radial coordinate $r$ for various combination of $(\alpha, B)$. Here, $m=1$}
    \label{fig:1}
\end{figure}

It is worth noted that the presence of string cloud alter the event horizon, the photon sphere, ISCO radii, horizon area and surface gravity in comparison to the Schwrazschild BH given in Table \ref{tab:1}.
\begin{table}[ht!]
\centering
\caption{\footnotesize Comparison of event horizon \((r_h)\), photon sphere radius \((r_{\rm ph})\), ISCO radius \((r_{\rm ISCO})\), horizon area \((\mathcal{A})\), and surface gravity \((\kappa)\) for Schwarzschild BH (SBH) and Letelier BH (LBH).}
\begin{tabular}{|c|c|c|c|c|c|}
\hline
BH & \(r_h\) & \(r_{\rm ph}\) & \(r_{\rm ISCO}\) & \(\mathcal{A}\) & \(\kappa\) \\[4pt]
\hline
SBH & $2m$ & $3m$ & $6m$ & $16 \pi m^2$ & $\dfrac{1}{4 m}$ \\
\hline
LBH & $\dfrac{2m}{1-\alpha}$ & $\dfrac{3m}{1-\alpha}$ & $\dfrac{6m}{1-\alpha}$ & $\dfrac{16\pi m^2}{(1-\alpha)^{2}}$ & $\dfrac{(1-\alpha)^2}{4 m}$ \\[4pt]
\hline
\end{tabular}
\label{tab:1}
\end{table}

{\bf Horizons:} For rotating BH space-time (\ref{bb1}), the horizon can be determined by setting $Q=0$, which implies $\Delta=0$. This yields (using $I_2=1-a^2 B^2$)
\begin{equation}
    r_{\pm}=\frac{m I_2\pm \sqrt{m^2 I_2-a^2 {\color{black} (1-\alpha)} I^2_1}}{(1-\alpha)\,I^2_1-B^2 m^2 I_2}\,I_1,\label{bb6}
\end{equation}
which are the outer and inner BH horizons. In the absence of string cloud, $\alpha=0$, the horizon Eq.~(\ref{bb6}) simplifies to
\begin{equation}
    r_{\pm}=\frac{m I_2\pm \sqrt{m^2 I_2-a^2 I^2_1}}{I^2_1-B^2 m^2 I_2}\,I_1\label{bb6a}
\end{equation}
which is similar to those expression reported in \cite{Podolsky2025}. Thus, we can conclude that the presence of string cloud in the current model alters the horizon radius.

\begin{figure}[ht!]
    \centering
    \includegraphics[width=0.8\linewidth]{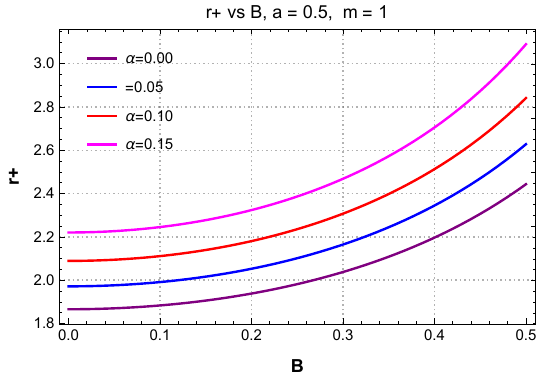}\\
    \includegraphics[width=0.8\linewidth]{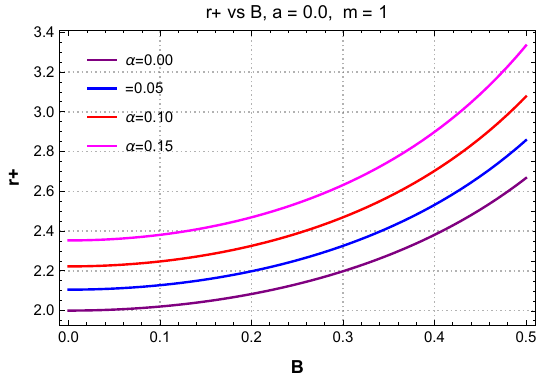}
    \caption{\footnotesize Behavior of the horizon radius $r_h=r_{+}$ using Eq.~(\ref{bb6}) as a function of the magnetic field strength $B$ with and without string clouds parameters $\alpha$.}
    \label{fig:2}
\end{figure}

For $B=0$, from Eq. (\ref{bb6}) one can recover $r_{\pm}=\frac{m \pm \sqrt{m^2-a^2 {\color{black} (1-\alpha)}}}{1-\alpha}$, valid for the Kerr-like BH surrounded by a cloud of strings which further reduces to Kerr horizon $r_{\pm}=m \pm \sqrt{m^2-a^2}$ in the absence of string cloud, $\alpha=0$. For zero rotation parameter, $a=0$, we get one horizon given by
\begin{equation}
    r_{+}=\frac{2 m}{1-\alpha-m^2 B^2}.\label{bb7}
\end{equation}
Notice that this horizon is located at greater values than the Letelier or Schwarzschild horizon. For $B=0$, one can recover Letelier horizon listed in Table \ref{tab:1}, and for $\alpha=0$, we recover Schwrazschild-BR horizon $r_{+}=2m/(1-m^2 B^2)$ reported in \cite{Podolsky2025}.

In the near extreme situation, arising when the discriminant of Eq.~(\ref{bb6}) vanishes, that is, 
\begin{equation}
    a^2 {\color{black} (1-\alpha)} I^2_1=m^2 I_2.\label{bb8}
\end{equation}
Substituting $I_1$ and $I_2$, we find the magnetic field value as follows:
\begin{equation}
    B^2_{\rm ext}=\frac{2 a^{-4}}{ {\color{black} (1-\alpha)}}\,\sqrt{m^2-a^2 {\color{black} (1-\alpha)}}\,(m-\sqrt{m^2-a^2 {\color{black} (1-\alpha)}}).\label{bb9}
\end{equation}
In the absence of string cloud, $\alpha=0$, this result similar to those reported in \cite{Podolsky2025} (see their Eq. (23)) that were obtained using the relation $a^2 I^2_1=m^2 I_2$. The presence of string cloud thus modified the extreme magnetic field value in comparison to the known expression. 

In the near extreme situation (\ref{bb8}), the two horizons from Eq. (\ref{bb6}) coincide at
\begin{equation}
    r_{\rm ext}=\frac{m}{(1-\alpha) I_1}\quad (\text{since } {\color{black} (1-\alpha)} I^2_1=m^2 I_2/a^2)\label{bb10}
\end{equation}
where
\begin{align}
    I_1&=1-\frac{a^{-2}}{{\color{black} (1-\alpha)}}\sqrt{m^2-a^2 {\color{black} (1-\alpha)}}\,(m-\sqrt{m^2-a^2 {\color{black} (1-\alpha)}}),\nonumber\\
    I_2&=1-\frac{2 a^{-2}}{{\color{black} (1-\alpha)}}\sqrt{m^2-a^2 {\color{black} (1-\alpha)}}\,(m-\sqrt{m^2-a^2 {\color{black} (1-\alpha)}}).
\end{align}
In the absence of string cloud, $\alpha =0$, the single horizon is similar to the ones reported in \cite{Podolsky2025}. For $a=m/{\color{black} \sqrt{1-\alpha}}$, we find $I_1=1=I_2$, and using Eq.~(\ref{bb10}) the horizon becomes \(r_{\rm ext}=m (1-\alpha)^{-1} >m\) which further simplifies to $r_{\rm ext}=m$ (exactly extremal Kerr horizon) in the absence of string cloud, $\alpha=0$. In other words, if string cloud is present in rotating BHs, the condition $a=m/{\color{black} \sqrt{1-\alpha}}$ characterizes a near extremal configuration rather than a truly extremal one, in contrast to the exactly extremal Kerr horizon obtained in the absence of string cloud.

{\bf Geodesic Motion}: We consider the geodesics motion in the equatorial plane defined by $\theta=\pi/2$. Moreover, for Schwrazschild-BR BH surrounded by a CoS given in the metric Eq. (\ref{bb3}), there exist two Killing vectors $\partial_{t}$ and $\partial_{\phi}$. Therefore, there are two conserved quantities $p_t=-\mathrm{E}$ and $p_{\phi}=\mathrm{L}$. Using $g_{\mu\nu}\,\dot{x}^{\mu}\,\dot{x}^{\nu}=0$ for photon motion, we get $\dot{r}^2=E^2-\frac{\mathrm{L}^2}{r^2}\,\mathcal{F}$. The photon sphere condition $\frac{d}{dr}\left(\frac{\mathcal{F}}{r^2}\right)=0$ which implies
\begin{equation}
    r_{\rm ph}=\frac{(1 - \alpha - B^{2}m^{2}) 
+ \sqrt{(1 - \alpha - B^{2}m^{2})^{2} - 12B^{2}m^{2}}}
{2B^{2}m}.\label{bb12}
\end{equation}
For $B=0$, we have $\mathcal{F} \to (1-\alpha-\tfrac{2 m}{r})$, and therefore, the photon sphere is listed in Table \ref{tab:1}.

To study photon trajectories, we determine the equation of orbits given by
\begin{equation}
    \left(\frac{1}{r^2}\frac{dr}{d\phi}\right)^2=\frac{\left[\frac{1}{\beta^2}-\left(B^2+\frac{1}{r^2}\right)\,\left(1-\alpha-m^2 B^2-\frac{2m}{r}\right)\right]}{(1+r^2 B^2)^2}.\label{trajectory-1}
\end{equation}
Transforming to a new variable via $u(\phi)=\frac{1}{r(\phi)}$ into the above relation yields
\begin{align}
    \left(\frac{du}{d\phi}\right)^2&=\frac{u^4}{(u^2+B^2)^2}\Big[2 m u^3-(1-\alpha-m^2 B^2)\,u^2+2 m B^2 u\nonumber\\
    &+\frac{1}{\beta^2}-B^2 (1-\alpha-m^2 B^2)\Big].\label{trajectory-2}
\end{align}
For zero magnetic field, $B=0$, we get $\left(\frac{du}{d\phi}\right)^2+(1-\alpha) u^2=2 m u^3+\frac{1}{\beta^2}$, where $\beta=\mathrm{L}/\mathrm{E}$, and its second-order equation after differentiating w. r. to $\phi$ simplifies as $\frac{d^2u}{d\phi^2}+(1-\alpha) u=3 m u^2$, the photon trajectory in Letelier-BH which reduces to the Schwarzschild case for $\alpha=0$.

For the motion of time-like particles of mass $m_0$, following the previous procedure and using $g_{\mu\nu}\,u^{\mu}\,u^{\nu}=-m_0$, we arrive at $\dot{r}^2=(1+B^2\,r^2)^2\,\left[\mathcal{E}^2-V_{\rm eff}(r)\right]$. Stable circular orbits
\(\dot{r}=0\) are possible only at the minimum of the effective potential  
\begin{equation}
    V_{\rm eff}(r)=\left[1-\alpha-m^2 B^2-\frac{2 m}{r}\right]\,\left[1+\mathcal{L}^2\,\left(\tfrac{1}{r^2}+B^2\right)\right].\label{bb13}
\end{equation}
Here $\mathcal{E}$ and $\mathcal{L}$, respectively are the energy and angular momentum of test particles per unit mass.

From the conditions $\mathcal{E}^{2} = V_{\mathrm{eff}}$,\quad 
$V'_{\mathrm{eff}}(r) = 0$,\quad and 
$V''_{\mathrm{eff}}(r)= 0$, where the prime denotes differentiation with respect to $r$, 
we obtain the expressions for the specific energy \((\mathcal{E}_{\rm specfic})\) and specific angular momentum \((\mathcal{L}_{\rm specific})\) of time-like particles per unit mass given by
\begin{align}
    \mathcal{E}_{\rm specfic}&=\frac{\left(1-\alpha-m^2 B^2-\frac{2 m}{r}\right)}{\sqrt{1-\alpha-m^2 B^2-\frac{3 m}{r}-m B^2 r}},\nonumber\\
    \mathcal{L}_{\rm specfic}&=\sqrt{\frac{m r}{1-\alpha-m^2 B^2-\frac{3 m}{r}-m B^2 r}}.\label{bb15}
\end{align}
And the ISCO is located at
\begin{align}
    r_{\rm ISCO}=\frac{6m}{1-\alpha-m^2 B^2}.\label{bb16}
\end{align}
For $\alpha=0$, corresponding to the absence of string cloud, the ISCO radius simplifies to $r_{\rm ISCO}=\frac{6m}{1-m^2 B^2}$ which is similar to the result reported in \cite{Podolsky2025} and Eq.~(\ref{bb15}) simplifies to
\begin{align}
    \mathcal{E}_{\rm specfic}&=\frac{\left(1-m^2 B^2-\frac{2 m}{r}\right)}{\sqrt{1-m^2 B^2-\frac{3 m}{r}-m B^2 r}},\nonumber\\
    \mathcal{L}_{\rm specfic}&=\sqrt{\frac{m r}{1-m^2 B^2-\frac{3 m}{r}-m B^2 r}}.\label{bb15a}
\end{align}
For zero magnetic field, $B=0$, the ISCO radius is listed in Table \ref{tab:1} and Eq.~(\ref{bb15}) simplifies to
\begin{align}
    \mathcal{E}_{\rm specfic}=\frac{\left(1-\alpha-\frac{2 m}{r}\right)}{\sqrt{1-\alpha-\frac{3 m}{r}}},\quad 
    \mathcal{L}_{\rm specfic}=\sqrt{\frac{m r}{1-\alpha-\frac{3 m}{r}}}.\label{bb15b}
\end{align}
The resulting ISCO is therefore located at larger radii than the ISCO in the Letelier or Schwarzschild spacetimes. Moreover, the specific energy and specific angular momentum of time-like particles orbiting at this circular orbit are influenced by the string parameter $\alpha$ and the magnetic field strength $B$, in contrast to the Schwarzschild BH case.

{\bf Thermodynamics}: Finally, we evaluate the basic thermodynamic quantities of the selected BHs, namely, the entropy $S$ and the temperature $T_H$ of their horizons. Using the well-known relations~\cite{Wald1984}, \(S = \frac{\mathcal{A}}{4}\), and \(T = \frac{\kappa}{2\pi},\) these quantities are determined by the horizon area $\mathcal{A}$ and the surface gravity $\kappa$, respectively. Before that, we determine the angular velocity of rotating metric in Eq.~(\ref{bb1}) given by
\begin{equation}
    \omega_{h}=-\frac{g_{t \phi}}{g_{\phi\phi}}\Big{|}_{r_{+}}=\frac{a}{r^2_{+}+a^2},\label{velocity}
\end{equation}
where $r_{+}$ is given in Eq.~(\ref{bb6}).

The horizon area is obtained by integrating over the angular coordinates of the metric~(\ref{bb1}) for fixed values of $t$ and $r = r_{+}$, namely, \(
\mathcal{A}(r_{+}) = \int_{0}^{2\pi C} \int_{0}^{\pi} \sqrt{g_{\theta\theta} g_{\phi\phi}} \, d\theta \, d\phi,\) where $C=[1+B^2(m^2 \tfrac{I_2}{I^2_1}-a^2)]^{-1}$ is the conicity parameter \cite{Podolsky2025}. We find  
\begin{equation}
\mathcal{A} = 2\pi C \,(r_{+}^2+a^2) \int_{0}^{\pi} \Omega^{-2}(r_{+}) \, \sin\theta \, d\theta=\frac{4 \pi C (r_{+}^2 + a^2)}{1 + B^2 r_{+}^2}, \label{bb17}    
\end{equation}
where $\Omega^2$ on the horizon $r_{+}$ is a constant: \(\Omega^2(r_{+}) = 1 + B^2 r_{+}^2.\) The entropy is $S=\mathcal{A}/4=\frac{\pi C (r_{+}^2 + a^2)}{1 + B^2 r_{+}^2}$. At this stage, it is worth noting that the horizon radius $r_{+}$ for $\alpha \neq 0$ is expressed in Eq.~(\ref{bb6}), while the corresponding expression for $\alpha = 0$ is given in Eq.~(\ref{bb6a}).

For non-rotating BHs, $a=0$, and hence $I_1=1=I_2$. The conicity parameter becomes $C=1/(1+m^2 B^2)$. The horizon area thus simplifies to
\begin{align}
\mathcal{A}(r_{+})=\frac{16 \pi m^2}{(1+m^2 B^2)\,[4 m^2 B^2+(1-\alpha-m^2 B^2)^2]},\label{bb18}    
\end{align}
Noted here that in the absence of string cloud, $\alpha =0$, the horizon area reduces to
\begin{align}
\mathcal{A}(r_{+})=16 \pi m^2 (1+m^2 B^2)^{-3}\label{bb19}    
\end{align}
which is similar to the expression reported in \cite{Podolsky2025}. Thus, we conclude that the presence of string cloud in static Schwrazschild-BR BH space-time (\ref{bb3}) modifies the horizon area in comparison to the known result. For $B=0$, the horizon area will be $\mathcal{A}(r_{+})=16 \pi m^2/(1-\alpha)^2$ which is similar to Letelier BH case listed in Table \ref{tab:1}.

The surface gravity $\kappa$ can be calculated using the formula \(\kappa=\frac{1}{2}\,Q'(r_{+})/(r^2_{+}+a^2)\). In our case at hand, we get 
\begin{align}
    \kappa&=\frac{1+B^2 r^2_{+}}{r^2_{+}+a^2}\,\frac{\Delta'(r_{+})}{2}=\frac{1}{2} \frac{1+B^2 r^2_{+}}{r^2_{+}+a^2}\,\left(2 \lambda\,r_{+}-2 m \frac{I_2}{I_1}\right)\nonumber\\
    &=
    \frac{1+B^2 r^2_{+}}{r^2_{+}+a^2}\,\sqrt{m^2 \frac{I^2_2}{I^2_1}-a^2 {\color{black} (1-\alpha)}+a^2 m^2 B^2 \frac{I_2}{I^2_1}},\label{bb20}
\end{align}
where we have used through Eq. (\ref{bb6}) the followings:
\begin{align}
    r_{+}+r_{-}&=\frac{2 m I_2/I_1}{\lambda},\quad \lambda=1-\alpha-m^2 B^2 \frac{I_2}{I^2_1},\nonumber\\
    r_{+}-r_{-}&=\frac{2}{\lambda}\,\sqrt{m^2 \frac{I^2_2}{I^2_1}-a^2{\color{black} (1-\alpha)}+a^2 m^2 B^2 \frac{I_2}{I^2_1}}.\label{bb20a}
\end{align}
For $B=0$, we have $I_1=1=I_2$, and hence, the surface gravity reduces to  $\kappa=\frac{\sqrt{m^2-a^2 {\color{black} (1-\alpha)}}}{r^2_{+}+a^2}$ with $r_{+}=\frac{m+\sqrt{m^2-a^2 {\color{black} (1-\alpha)}}}{1-\alpha}$ which further reduces to Kerr horizon $r_{+}=m+\sqrt{m^2-a^2}$ in the absence of string cloud, $\alpha=0$. At this stage, it is worth mentioning that surface gravity expression given in Eq.~(\ref{bb20}) for $\alpha=0$ case coincides with those result reported in \cite{Podolsky2025}. Thus, we can conclude that the presence of string cloud in the current model alter the surface gravity, and consequently, the temperature in comparison to the known result.

\begin{figure}[ht!]
    \centering
    \includegraphics[width=0.8\linewidth]{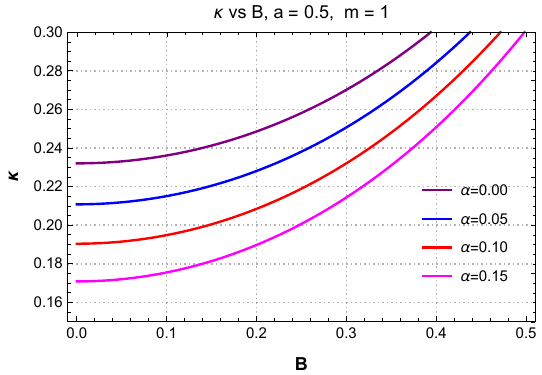}\\
    \includegraphics[width=0.8\linewidth]{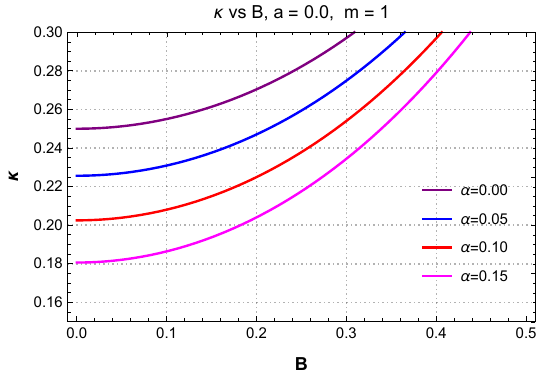}
    \caption{\footnotesize Behavior of the surface gravity $\kappa$ for non-rotating and rotating BHs as a function of the magnetic field $B$ with and without string cloud parameter $\alpha$.}
    \label{fig:3}
\end{figure}

Figure \ref{fig:3} illustrates the variation of the surface gravity $\kappa$ with the magnetic field strength $B$ for both rotating and non-rotating BHs, with and without the string parameter $\alpha$. For instance, at $B = 0.1$ and $\alpha=0$, the rotating BH exhibits a lower surface gravity than the non-rotating one. The inclusion of a string cloud parameter $\alpha$ further suppresses $\kappa$ in both cases. Conversely, as shown in Fig. \ref{fig:2}, the string parameter $\alpha$ increases the horizon radius for both rotating and non-rotating configurations.

For non-rotating BH, $a=0$, we have $I_1=1=I_2$, and horizon $r_{+}$ is given in Eq.~(\ref{bb7}). Therefore, the surface gravity of BH metric (\ref{bb5}) becomes 
\begin{equation}
\kappa=m\,(B^2+1/r^2_{+})=\frac{4 m^2 B^2+(1-\alpha-m^2 B^2)^2}{4 m}.\label{bb20b} 
\end{equation}
In the absence of string cloud, $\alpha=0$, the surface gravity reduces to those result reported in \cite{Podolsky2025}. Moreover, in the absence of magnetic field $B=0$, the surface gravity is listed in Table \ref{tab:1}.

The Hawking temperature is given by ($ a \neq 0$)
\begin{equation}
    T=\frac{\kappa}{2 \pi}=\frac{1}{2\pi}\frac{1+B^2 r^2_{+}}{r^2_{+}+a^2}\,\sqrt{m^2 \frac{I^2_2}{I^2_1}-a^2 {\color{black} (1-\alpha)}+a^2 m^2 B^2 \frac{I_2}{I^2_1}}.\label{bb21}
\end{equation}
From the above analysis, one can see that the string parameter $\alpha$ has the effects to angular velocity ($\omega_h$), horizon area ($\mathcal{A}$), and temperature  $(T)$ by affecting the horizon radius $r_{+}$. For extreme BH with a degenerate horizon at $r_{+}=r_{\rm ext}$ given in Eq.~(\ref{bb10}), the temperature reduces to
\begin{equation}
    T=\frac{1}{2\pi}\frac{1+B^2_{\rm ext} r^2_{\rm ext}}{r^2_{\rm ext}+a^2}\,\sqrt{m^2 \frac{I^2_2}{I^2_1}-a^2 {\color{black} (1-\alpha)}+a^2 m^2 B^2_{\rm ext} \frac{I_2}{I^2_1}},\label{bb22}
\end{equation}
where $B_{\rm ext}$ is given in Eq.~(\ref{bb9}). In the extremal case where \( a = m/{\color{black} \sqrt{1-\alpha}} \), we find \( I_1 = 1=I_2 \), \( B_{\rm ext} = 0 \), and \( r_{\rm ext} = m (1 - \alpha)^{-1} \). Consequently, the temperature vanishes, \( T = 0 \).

By combining Eqs.~(\ref{bb17}) and (\ref{bb20}), we find the following relation
\begin{equation}
    2TS=\frac{\mathcal{A}(r_{+}) \kappa}{4\pi}=\frac{\sqrt{m^2 \frac{I^2_2}{I^2_1}-a^2 {\color{black} (1-\alpha)}+a^2 m^2 B^2 \frac{I_2}{I^2_1}}}{\left[1+B^2\left(m^2 \frac{I_2}{I^2_1}-a^2\right)\right]}.\label{bb23}
\end{equation}

For $a=0$ which results $I_1=1=I_2$, we find
\begin{align}
    T&=\frac{m}{2\pi}\,\frac{1+B^2 r^2_{+}}{r^2_{+}},\nonumber\\
    S&=\pi C \frac{r^2_{+}}{1+B^2 r^2_{+}},\quad r_{+}=\frac{2 m}{1-\alpha-m^2 B^2},\nonumber\\
    2 T S&=mC.\label{bb24}
\end{align}
This represents a modified Smarr relation \cite{Smarr1973} for a magnetized static BH surrounded by a string cloud (\ref{bb3}), with the conicity parameter $C=(1+m^2 B^2)^{-1}$ as defined previously. The first law of thermodynamics takes the form
\begin{equation}
    dm=T dS +\Phi_{\alpha}\,d\alpha+\Phi_{B} dB,\label{bb25}
\end{equation}
where the conjugate potentials are defined as
\begin{align}
    \Phi_{\alpha}&=\left(\frac{\partial m}{\partial \alpha}\right)_{S,B},\quad 
    \Phi_B=\left(\frac{\partial m}{\partial B}\right)_{S,\alpha}.\label{bb26}
\end{align}

For $B=0$ which results $I_1=1=I_2$, we find
\begin{align}
    T&=\frac{1}{2 \pi}\,\frac{1}{r^2_{+}+a^2}\,\sqrt{m^2-a^2 (1-\alpha)},\nonumber\\
    S&=\pi (r^2_{+}+a^2),\quad r_{+}=\frac{m+\sqrt{m^2-a^2 (1-\alpha)}}{1-\alpha},\nonumber\\
    2 T S&=\sqrt{m^2-a^2 (1-\alpha)}=M_{\rm irr}, \label{bb27}
\end{align}
where $M_{\rm irr}$ is the irreducible mass, which is similar to the Smarr relation for Kerr-BH \cite{Smarr1973} and reduces to the Letelier BH case for $a=0$. The first law of thermodynamics is given by
\begin{equation}
    dm=T dS+\Omega_h dJ +\Phi_{\alpha}\,d\alpha,\label{bb28}
\end{equation}
where $J=am$ denotes the angular momentum, $\Omega_h=\omega_h=a/(r^2_{+}+a^2)$ is the angular velocity at the horizon, and $\Phi_{\alpha}=(\partial m/\partial \alpha)_{S,J}$ is the thermodynamic conjugate potential associated with the string parameter. This formulation reveals how the string cloud modifies the standard Kerr thermodynamics by introducing an additional contribution $\Phi_{\alpha}\,d\alpha$ that accounts for variations in the string density surrounding the BH.

\section{Conclusion}

Magnetic fields play a crucial role in the physics of black holes, especially in astrophysical contexts. They influence the dynamics of accretion disks, govern the formation and collimation of relativistic jets, and facilitate energy extraction mechanisms such as the Blandford-Znajek process. Strong magnetic fields can also modify the spacetime structure around rotating black holes, affecting particle trajectories, radiation emission, and observable signatures like black hole shadows. On the other hand, cosmic strings, an one-dimensional topological defects predicted by certain grand unified theories-can have significant gravitational effects when present near black holes. They alter geodesic structure and thermodynamic properties. Studying the combined effects of magnetic fields and cosmic strings provides deeper insights into exotic astrophysical environments and offers potential avenues to test fundamental physics beyond general relativity.   

The Event Horizon Telescope (EHT) has recently imaged the polarized emission around the supermassive black hole (SMBH) M87* on event-horizon scales, estimating the average magnetic field strength to be approximately 1-30 G \cite{Akiyama2021EHT}. In another paper \cite{Akiyama2023EHT}, they reported results showing how light from the edge of the supermassive black hole M87* spirals as it escapes the intense gravitational pull, producing a phenomenon known as circular polarization. The preferred rotation of light’s electric field-clockwise or counterclockwise-carries valuable information about the magnetic field structure and the types of high-energy particles present around the black hole. These new observations reinforce earlier EHT findings, indicating that the magnetic field near M87* is strong enough to occasionally hinder the black hole from accreting nearby matter.

In the present study, we have systematically investigated the effects of both an external Bertotti-Robinson uniform magnetic field and surrounding string clouds on the properties of black holes, considering both rotating and non-rotating configurations. Our analysis demonstrated that these external factors significantly alter the horizon structure, leading to measurable changes in the horizon radius compared to the standard Kerr and Schwarzschild solutions. As a direct consequence of the modified horizon radius, key quantities such as the horizon area, black hole entropy, and Hawking temperature also exhibit corresponding variations. Beyond the horizon properties, we find that the photon sphere, the trajectories of photons, and the innermost stable circular orbit (ISCO) are all sensitive to the presence of the magnetic field and string clouds, indicating that the geodesic motion of test particles is strongly influenced by these external parameters. Notably, the inclusion of string clouds introduces additional modifications that are absent in the black hole solutions \cite{Podolsky2025}, affecting both the horizon geometry, consequently, the thermodynamic properties and the dynamics of test particles in the surrounding spacetime. These findings show that the interplay between magnetic fields and string clouds can lead to rich phenomenology, resulting in deviations from previously reported results, such as those in Ref. \cite{Podolsky2025}, and underscore the importance of considering such environmental effects in realistic black hole models.

\section*{Acknowledgments}

F.A. is grateful to the Inter University Centre for Astronomy and Astrophysics (IUCAA), Pune, India, for the opportunity to hold a visiting associateship. \.{I}.~S. extends appreciation to T\"{U}B\.{I}TAK, and ANKOS for their assistance. Additionally, he acknowledges the support from COST Actions CA22113, CA21106, CA23130, and CA23115, which have been pivotal in enhancing networking efforts.

\section*{Funding statement}

No fund has received for this article.

\section*{Data Availability Statement}

No data were generated or analyzed in this article.

\bibliography{apssamp} 

@article{Cardoso:2004uz,
    author = {Cardoso, Vitor and Dias, Oscar J. C. and Lemos, Jose P. S.},
    journal = "Phys. Rev. D",
    volume = "70",
    pages = "024002",
    year = "2004",
    doi = "10.1103/PhysRevD.70.024002",
}

@article{Mazharimousavi:2009vh,
    author = "Mazharimousavi, S. Habib and Halilsoy, M. and Sakalli, I. and Gurtug, O.",
    doi = "10.1088/0264-9381/27/10/105005",
    journal = "Class. Quant. Grav.",
    volume = "27",
    pages = "105005",
    year = "2010"
}

@article{Al-Badawi:2008ucc,
    author = "Al-Badawi, A. and Sakalli, I.",
    doi = "10.1063/1.2912725",
    journal = "J. Math. Phys.",
    volume = "49",
    pages = "052501",
    year = "2008"
}

@article{Badawi32,
    author = "Al-Badawi, A. and Halilsoy, M.",
    doi = "10.1063/1.2912725",
    journal = "IL Nuovo Cimento ",
    volume = "119B",
    pages = "931",
    year = "2004"
}

@article{Podolsky2025,
  author    = {J. Podolsky and H. Ovcharenko},
  journal   = {Phys. Rev. Lett.},
  year      = {2025},
  volume    = {135},
  pages     = {181401},
  doi       = {10.1103/rfgv-ybz5}
}

@book{Misner1973,
  author    = {C. W. Misner and K. S. Thorne and J. A. Wheeler},
  title     = {Gravitation},
  publisher = {W. H. Freeman and Company},
  year      = {1973},
  address   = {San Francisco}
}

@book{Carroll2004,
  author    = {S. M. Carroll},
  title     = {Spacetime and Geometry: An Introduction to General Relativity},
  publisher   = {Addison-Wesley},
  year      = {2004}
}

@article{Kerr1963,
  author    = {R. P. Kerr},
  journal   = {Phys. Rev. Lett.},
  year      = {1963},
  volume    = {11},
  pages     = {237--238},
  doi       = {10.1103/PhysRevLett.11.237}
}

@article{Abbott2016GW,
  author       = {B. P. Abbott and  R. Abbott and et al. [LIGO Scientific Collaboration and the Virgo Collaboration]},
  journal      = {Phys. Rev. Lett.},
  year         = {2016},
  volume       = {116},
  pages        = {061102},
  doi          = {10.1103/PhysRevLett.116.061102},
}

@article{Akiyama2019EHT,
  author    = {K. Akiyama and et al. [Event Horizon Telescope Collaboration]},
  journal   = {ApJL},
  year      = {2019},
  volume    = {875},
  pages     = {L1},
  doi       = {10.3847/2041-8213/ab0ec7}
}

@article{Akiyama2019,
  author    = {K. Akiyama and et al. [Event Horizon Telescope Collaboration]},
  journal   = {ApJL},
  year      = {2019},
  volume    = {875},
  pages     = {L4},
  doi       = {10.3847/2041-8213/ab0e85}
}

@book{Kiritsis2007,
  author    = {Elias Kiritsis},
  title     = {String Theory in a Nutshell},
  publisher = {Princeton University Press},
  address   = {Princeton, NJ},
  year      = {2007},
  isbn      = {978-0-691-12230-4}
}

@article{Letelier1979,
  author  = {P. S. Letelier},
  journal = {Phys. Rev. D},
  volume  = {20},
  pages   = {1294},
  year    = {1979},
  doi     = {https://doi.org/10.1103/PhysRevD.20.1294}
}

@article{Toledo2020,
  author    = {J. M. Toledo and V. B. Bezerra},
  journal   = {Gen. Relativ. Gravit.},
  year      = {2020},
  volume    = {52},
  pages     = {34},
  doi       = {10.1007/s10714-020-02683-1},
}

@article{Newman1965Kerr,
  author    = {E. T. Newman and A. I. Janis},
  journal   = {J. Math. Phys.},
  year      = {1965},
  volume    = {6},
  pages     = {915--917},
  doi       = {10.1063/1.1704350},
}

@article{Newman1965Rotating,
  author    = {E. T. Newman and E. Couch and K. Chinnapared and A. Exton and A. Prakash and R. Torrence},
  journal   = {J. Math. Phys.},
  year      = {1965},
  volume    = {6},
  pages     = {918--919},
  doi       = {10.1063/1.1704351},
}

@article{AzregAinou2014,
  author    = {Mustapha Azreg-Aïnou},
  journal   = {Phys. Rev. D},
  year      = {2014},
  volume    = {90},
  pages     = {064041},
  doi       = {10.1103/PhysRevD.90.064041},
}

@article{Li2021,
  author  = {Z. Li and T. Zhou},
  journal = {Phys. Rev. D},
  volume  = {104},
  pages   = {104044},
  year    = {2021},
  doi      = {10.1103/PhysRevD.104.104044}
}

@article{Sun2024,
  author    = {Qi Sun and Yu Zhang and Chen-Hao Xie and Qi-Quan Li},
  journal   = {Phys. Dark Univ.},
  year      = {2024},
  volume    = {46},
  pages     = {101599},
  issn      = {2212-6864},
  doi       = {10.1016/j.dark.2024.101599},
}

@article{Blandford1977,
  author    = {R. D. Blandford and R. L. Znajek},
  journal   = {MNRAS},
  volume    = {179},
  pages     = {433--456},
  year      = {1977},
  doi       = {10.1093/mnras/179.3.433}
}

@article{Meier2001,
  author    = {D. L. Meier},
  journal   = {ApJ},
  volume    = {548},
  pages     = {L9--L12},
  year      = {2001},
  doi       = {10.1086/318921}
}

@article{Janiuk2022,
  author    = {Agnieszka Janiuk and Bestin James},
  journal   = {A\&A},
  volume    = {668},
  pages     = {A66},
  year      = {2022},
  doi       = {10.1051/0004-6361/202244196}
}

@article{Ressler2025,
    author   = {Sean M. Ressler and Luciano Combi and Bart Ripperda and Elias R. Most},
    journal  = {ApJL},
    year     = {2025},
    volume   = {979},
    pages    = {L24},
    doi      = {10.3847/2041-8213/ad9eb5}
}

@article{Bertotti1959,
  author    = {B. Bertotti},
  journal   = {Phys. Rev.},
  year      = {1959},
  volume    = {116},
  pages     = {1331--1333},
  doi       = {10.1103/PhysRev.116.1331}
}

@article{Robinson1961,
  author    = {I. Robinson},
  journal   = {J. Math. Phys.},
  year      = {1961},
  volume    = {2},
  pages     = {290},
 doi        ={10.1063/1.1703712}
}

@article{Asher1960,
    author    = {Asher Peres},
    journal   = {Phys. Rev.},
    year      = {1960},
    volume    ={118},
    pages     ={1105},
    doi       ={10.1103/PhysRev.118.1105}
}

@article{Zeng2025,
  author       = {Xiao-Xiong Zeng and Chen-Yu Yang and Hao Yu},
  journal      = {Eur. Phys. J C},
  volume       = {85},
  pages        = {1242},
  year         = {2025},
  doi          = {10.1140/epjc/s10052-025-14989-y},
}

@book{Wald1984,
  author    = {R. M. Wald},
  title     = {General Relativity},
  publisher = {University of Chicago Press},
  address   = {Chicago and London},
  year      = {1984}
}

@article{Smarr1973,
  author       = {L. Smarr},
  title        = {Mass formula for Kerr black holes},
  journal      = {Phys. Rev. Lett.},
  volume       = {30},
  number       = {2},
  pages        = {71--73},
  year         = {1973},
  doi          = {10.1103/PhysRevLett.30.71}
}

@article{Akiyama2021EHT,
  author       = {K. Akiyama and et al. [The Event Horizon Telescope Collaboration]},
  journal      = {ApJL},
  volume       = {910},
  pages        = {L13},
  year         = {2021},
  doi          = {10.3847/2041-8213/abe4de}
}

@article{Akiyama2023EHT,
  author       = {K. Akiyama and et al. [The Event Horizon Telescope Collaboration]},
  journal      = {ApJL},
  volume       = {957},
  pages        = {L20},
  year         = {2023},
  doi          = {10.3847/2041-8213/ace11b}
}

\end{document}